# Text Steganography using LSB insertion method along with Chaos theory

Bhavana.S[1] and K.L.Sudha[2]

[1]Student (MTech), Department of ECE, DSCE, K.S.Layout, Bangalore
`bhavana.bhs2gmail.com`

[2]Professor, Department of ECE, DSCE, K.S.Layout, Bangalore
`klsudha1@rediffmail.com`

*ABSTRACT*

*The art of information hiding has been around nearly as long as the need for covert communication. Steganography, the concealing of information, arose early on as an extremely useful method for covert information transmission. Steganography is the art of hiding secret message within a larger image or message such that the hidden message or an image is undetectable; this is in contrast to cryptography, where the existence of the message itself is not disguised, but the content is obscure. The goal of a steganographic method is to minimize the visually apparent and statistical differences between the cover data and a steganogram while maximizing the size of the payload. Current digital image steganography presents the challenge of hiding message in a digital image in a way that is robust to image manipulation and attack. This paper explains about how a secret message can be hidden into an image using least significant bit insertion method along with chaos.*

## 1. INTRODUCTION

Chaos word has been derived from the Greek, which refers to unpredictability and it is defined as a study of nonlinear dynamic system. Chaos theory is a mathematical physics which was developed by Edward Lopez. Chaos is suitable for steganography, as it is closely related to some dynamics of its own characteristics. The behavior of the chaos system, under certain conditions, presents phenomena which are characterized by sensitivities to initial conditions and system parameters. Through the sensitivities, the system responses act to be random [1]. The main advantages of the chaotic steganographical approach include: Easy implementation, more randomness, sensitivity to initial conditions, non-periodic, and confidential.

Steganography is a process of hiding a secret message into an image or hiding a secret image into a cover image. There are two basic methods implemented in steganography: Least significant bit (LSB) Spatial Domain Technique and Transform-based (DCT) - Frequency Domain Technique. LSB steganography is a one of the simplest methods. Data hidden in images using this method is highly sensitive to image alteration & vulnerable to attack. DCT steganography is potentially more resistant to loss from image manipulation and increases the difficulty to a potential attacker. In this paper, we mainly deal with only LSB steganography method.

For hiding secret message in a cover image, henon map is being used. In the henon map, chaotic behavior can arise from very simple non-linear dynamical equations. It is a 2D map;

Hence it has two initial conditions. The map was introduced by Michel Hénon. Mathematically, the henon map is written:

x(n+1) = 1-a*x(n)^2+y(n);  & Y(n +1) = b*x(n);

Xn is a number between zero and one, a & b represents the initial conditions, where in we get maximum randomness. A rough description of chaos is that chaotic systems exhibit a great sensitivity to initial conditions—a property of the henon map for most values of a is about 1.4 & above and b is 0.1. In the present paper a=1.5, and b=0.1 is chosen. A common source of such sensitivity to initial conditions is that the map represents a repeated folding and stretching of the space on which it is defined. The bifurcation diagram for the henon map is shown in Figure 1.

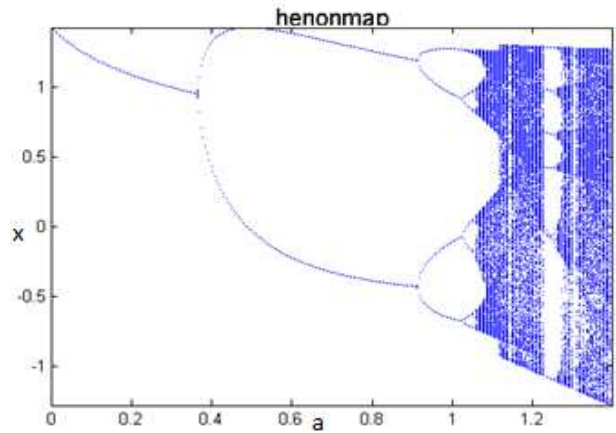

Figure 1. Bifurcation diagram for Henon map

A large number of applications in real systems, both man-made and natural, are being investigated using this novel approach of nonlinear dynamics. Many chaos-based steganographic methods have been presented and discussed in the last few decades. In paper [2], authors have focused on the developing the techniques that can help hiding messages on the basis of random numbers logic. They have concentrated upon using Least Significant Bit conversion but is not limited to it. It has involved other methods for steganography discussed in paper. This paper is an effort to explore the real power of random numbers to hide the messages in secure and customized way. In paper [3], they have proposed the use of encryption using Arnold and 2D Logistic methods. The chaotically encrypted data is embedded using two steganographic methods. A steganography method is proposed in paper [4] to embed information within an encrypted image data randomly. This paper offers a simple and strong way to conceal the data in the encrypted image. Thus, it will be used to reduce the chance of the encrypted image being detected and then enhance the security level of the encrypted images. The paper [5] presents a large capacity of steganography algorithm, and it can embed the secret image adaptively into the still image based on chaotic mapping and human visual characteristics. Simulation results show that it has high capacity, a good invisibility, and that it is robust for the common image processing like JPEG compression, cropping etc.

## 2. PROPOSED STEGANOGRAPHIC ALGORITHM

### 2.1 Hiding a secret message into another image

Consider a gray scale image (Io) with dimension M×N; M, N represents rows and column of intensity level. Take the message which is to be embedded. For example the message to be embedded is "Inception is a movie on dreams". Each letter in the message is converted to ASCII codes. Consider henon map which produces chaotic sequence, which is generated by using the equation given below:

x(n+1)=1-a*x(n)^2+y(n);  &  y(n+1)=b*x(n);

We have used the following values for the constants 'a' and 'b' to get a random sequence:

a=1.5 & b=0.1

The sequences are converted to binary by taking their average as threshold value. Each bit of the converted message is XORed with this chaotic sequence. Each XORed bit is again XORed with the least significant bit of the pixel of the image selected as the cover image. For example Consider, the binary value of I (ASCII 73), 01001001 (a) and it is XORed with the binary chaotic sequence, say 10010001000100…….(b), which is obtained using threshold value. After XORing we get the output as 11011000…..(c). Each of these bits is now XORed with LSB of individual pixel of the cover image. Say the 1st pixel of the image is 56 which have binary value 00111000 (d). The LSB is '0'. It is XORed with '1', which is obtained in the above sequence. The same is repeated until all the bits are embedded in the image.   Now, the stego image is obtained. Here, the secret message what is chosen is of length 30. The cover image is pokemon image with dimension 80*80.

The desteganography is done to get back the secret message by following the reverse process followed for steganography. As we know that each pixel of steganography image is of 8 bits, the first 7 bits are masked to obtain the sequence (c). Generate the chaotic sequence (b) with same initial conditions used for steganography. The sequence of bits obtained in (c) is XORed with (b) to retrieve the secret message.

All the above results are obtained from the simulation of steganography algorithm in Matlab Simulator.

## 3. PERFORMANCE ANALYSIS

The phrase peak signal-to-noise ratio, often abbreviated PSNR, is an engineering terminology that defines the ratio between the maximum possible power of a signal and the power of corrupting noise that affects the representation of the signal.

The PSNR is most often used as an important parameter to calibrate the quality of reconstruction of steganographic images. The signal in this case is the original image, and the noise is the error introduced by some steganography algorithm. It is most easily defined via the mean squared error (MSE) which for two m×n monochrome images I and K where one of the images is considered a noisy approximation of the other is defined as:

$$\text{MSE} = \frac{1}{mn} \sum_{i=0}^{m-1} \sum_{j=0}^{n-1} [i(i,j) - K(i,j)]2$$

Where, I=Original image.
       K=Restored image.
       m,n=no of pixels

$$\text{PSNR} = 10\log_{10}(\frac{MAX}{\sqrt{MSE}})$$

Where, MAX=255, for 8-bit image.

Table 1. PSNR value

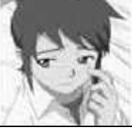

## 4. EXPERIMENTAL RESULTS

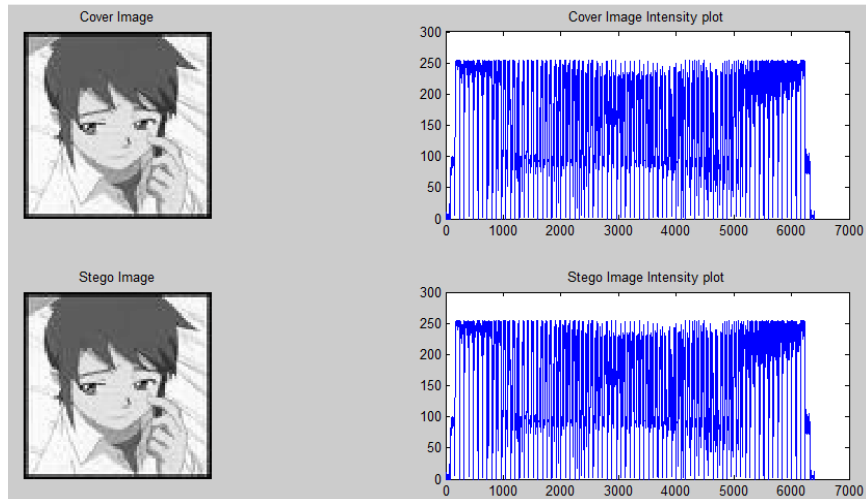

Figure 2. Cover image, Stego image and their intensity plots

Restored message is "Inception is a movie on dreams".

## 5. CONCLUSION

This paper describes about a steganography techniques, for hiding text message using the concept of non-linear dynamic system (chaos). The chaos system is highly sensitive to initial values and parameters of the system. The proposed algorithm provides security and maintains secrecy of the secret message and provides more randomness since we are using chaos which is sensitivity to initial conditions. The performance analysis is being done. The PSNR value is also being calculated.

## ACKNOWLEDGEMENTS

The work described in this paper is supported by the ISRO (Sanction order NO. E.33011/74/2011-V).

## REFERENCES

[1] Sanjeev Manchanda, Mayank Dave, and S. B. Singh, "Customized and secure image Steganography through Random numbers logic", *Signal Processing: An international Journal*, Vol. 1: issue.

[2] K. Ganesan, B. Venkatalakshmi, and R. Krishna Moothy, (2004), "Steganography using enhanced chaotic encryption technique", Available:http://*www.niitcrcs.com/iccs/iccs2004*/Papers/145%20B%20Venkatalakshmi.pdf.

[3] Dr. K. L. Sudha, and Manjunath Prasad, (Aug. 2011)," Chaos image encryption using pixel shuffling with henon map," *Elixir Elec. Engg*. 38, pp 4492-4495.

[4]     Mohammad Ali Bani Younes, and Aman Jantan, (June 2008 ), "A new steganography approach for image encryption exchange by using least significant bit insertion", *IJCSNS International Journal of Computer Science and Network Security*, Vol. 8 No.6.

[5]     Peipei Liu, Zhongliang Zhu, Hongxia Wang, and Tianyum Yan, "A novel image steganography Using chaotic map and visual model", Available: *www.atlantis press.com/php/download_paper.php*?id=1452

[6]     Jayakar.T, Tobin.C, Madhavi.K, Murali.K, "Chaos based spread spectrum image steganography", *IEEE Journal*, *volume-50, issue-2, On Page(s): 587 – 590.*

[7]     Lifang Yu, Yao Zhao, Rongrong Ni, Ting Li, "Improved adaptive LSB Steganography based on chaos and genetic algorithm",*EURASIP Journal on Advances in Signal Processing (2010) Volume: 2010, Publisher: Hindawi*

**Authors**

Bhavana.S, presently is a MTech student in the     branch Digital electronics and communication, Dept. Of ECE, Dayananda Sagar College of Engineering, Bangalore, India. She obtained her Bachelor's degree in electronics and communication engineering from Nagarjuna  College of engineering & technology, Bangalore, India and pursuing Masters  from Dayananda Sagar College of Engineering, Bangalore, India.

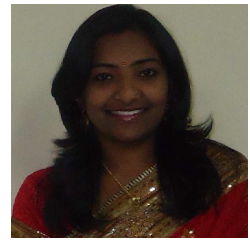

Dr.K.L.Sudha, presently working as professor in ECE department, Dayananda Sagar College of Engineering, Bangalore, India has 16 years of teaching experience in Engineering Colleges. She obtained her Bachelor's degree in electronics engineering from Mysore University and Masters from Bangalore University. She got Ph.D for her work on "Detection of FH CDMA signals in time varying channel" from Osmania University, Hyderabad.  She has published more than 15 research papers in national / International journals and conferences. Her research interests are in Wireless communication, coding theory, image processing and chaotic theory.

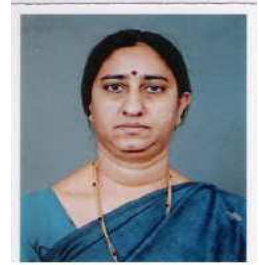